\documentclass[onecolumn,amsmath,amssymb]{qm2008_abs}
\usepackage{graphicx}
\usepackage{dcolumn}
\usepackage{bm}
\begin{document}

\title{{Investigation of high p$_{t}$ events in Nucleus-Nucleus collisions using the Hijing event generator }}

\bigskip
\bigskip
\author{\large Natasha Sharma and  Madan M. Aggarwal}
\email{snatasha.pu@gmail.com}
\affiliation{Panjab University, Chandigarh, India}
\bigskip
\bigskip

\begin{abstract}
   In recent years lot of interest has been observed in the
   nucleus-nucleus collisions at RHIC energies in phenomena
   related to high $p_{t}$ physics~\cite{ref1}. The suppression of
   high $p_{t}$ particles and disappearance of back-to-back jets
   compared to the scaling with number of binary nucleon-nucleon collisions 
indicates that a nearly perfect liquid is produced
   in these collisions. Results on self shadowing of high $p_{t}$ events
   are presented using hadron multiplicity associated to high $p_{t}$
   and  unbiased events in nucleus-nucleus collisions~\cite{ref2} obtained
   from the hijing event generator. 
\end{abstract}

\maketitle

\section{Introduction and Theory}

It has been shown by Cunqueiro, Deus and Pajares \cite{ref2} that the difference
between the multiplicity associated to high $p_{t}$ events and unbiased
multiplicity is given by the normalised variance of the unbiased multiplicity
indicating thereby the self-shadowing of the high $p_{t}$ events.
We here reproduce some of the equations of ref.\cite{ref2} for the sake
of clarity. In hadron-nucleus collisions, the inelastic unbiased cross
section is defined as:
\begin{eqnarray}
\sigma^{hA}(b)= {\sum\limits_{n=1}^{A}}{\binom{A}{n}}(\sigma T(b))^{n}(1-\sigma T(b))^{A-n}
\end{eqnarray}
where $\sigma$T(b), the collision probability,  is further divided into
two classes viz., a collision giving rise to high $p_{t}$ particle termed
as events of type C with cross-section $\sigma_{C}$ and rest of the events
without high $p_{t}$ particle i.e., non-C type events with cross-section 
$\sigma_{NC}$. Thus :

\begin{eqnarray}
(\sigma T(b))^{n}={\sum\limits_{i=0}^{n}}{\binom{A}{n}}(\sigma_{C})^{i}(\sigma_{NC})^{n-i}T(b)^{n},
\end{eqnarray}

 The final cross section for events of type C must contain at least one elementary $\sigma_{C}$ in the sum~\cite{ref3} i.e.,
\begin{eqnarray}
\sigma^{hA}_{C}={\sum\limits_{n=1}^{A}} {\sum\limits_{i=1}^{n}} \sigma^{i}_{C} \sigma^{n-i}_{NC} T(b)^{n} X (1-(\sigma_{C}+\sigma_{NC}) T(b)^{A-n}
\end{eqnarray}

\begin{eqnarray}
=1-(1-\sigma_{C}T(b))^{A}
\end{eqnarray}
This equation shows that C-events are self-shadowed, in the sense that their cross section depends only on their cross section in nucleon-nucleon collision.
This is also true for nucleus-nucleus collisions \cite{ref4, ref5}.
Taking $\alpha_{C}$ as the probability for an elementary collision to be of
 type C, N($\nu$) total number of events, and N$_{C}$($\nu$)
 total number of events of type C, 
 the probability distribution for C type events with $\nu$ collisions in the
limit of small $\alpha_{c}$ can be written as \cite{ref2}:
\begin{eqnarray}
P_{C}(\nu)=\frac{\alpha_{C}\nu N(\nu)}{\sum_{\nu}N_{C}(\nu)}=\frac{\nu N(\nu)}{<\nu> \sum_{\nu} N(\nu)}=\frac{\nu P(\nu)}{<\nu>}
\end{eqnarray}
It was proposed~\cite{ref2} that if the total multiplicity P(n) is obtained by the 
convolution of the
elementary multiplicity distributions p(n), 
the total dispersion D is related to the dispersion d and multiplicity $\bar n$
, of the distribution of elementary interaction as
\begin{eqnarray}
\frac{D^{2}}{<n>^{2}}=\frac{<\nu^{2}>-<\nu>^{2}}{<\nu>^{2}} + \frac{d^{2}}{<\nu>\bar n^{2}}
\end{eqnarray}
As in nucleus-nucleus collision $\nu$ is very high so neglecting IInd term
one gets :
\begin{eqnarray}
\frac{D^{2}}{<n>^{2}}=\frac{<\nu^{2}>-<\nu>^{2}}{<\nu>^{2}}
\end{eqnarray}
Hence, normalized dispersion of the total multiplicity is approximated by
the normalized dispersion of the number of elementary interactions. This
argument is
used to extend Eq. 5  to the multiplicity distribution~\cite{ref2} i.e.,
\begin{eqnarray}
P_{C}(n)  =  \frac{nP(n)}{<n>}
\end{eqnarray}
which can be written as :
\begin{eqnarray}
<n>_{C} - <n> = \frac{D^{2}}{<n>}
\end{eqnarray}
Therefore, the difference between the average multiplicity associated with 
high p$_{t}$ events and unbiased average multiplicity is given by 
the normalized variance of the unbiased multiplicity distribution if high
$p_{t}$ events are self-shadowed.

\section{Results and discussion}
We generated one million Au+Au minimum bias events at $\sqrt{s_{NN}} =$ 200 GeV and 100K
Pb+Pb  minimum bias events at $\sqrt{s_{NN}} =$ 11 TeV using
Hijing Event generator with default setting. The analysis was done
in the  range -1.0 $<$ y $<$ 1.0 and $0^{o}$ $\le$
 $\phi$ $\le$ $360^{o}$ for different $p_{t}$ cuts. The centrality bins were
calculated using the number of participants in a collision. We used eight
equal spacing centrality bins of $N_{part}$ i.e., 0-50, 50-100, 150-200,
200-250, 250-300, 300-350, and 350-400 representing, respectively, as centralities
1, 2, 3, 4, 5, 6, 7, and 8 for Au+Au collisions. In case of Pb+Pb collisions
we also used the centrality
bin corresponding to 400 $<$ $N_{part}$ $< $ 450, representing the 9th centrality
bin. Events having at least one high p$_{t}$ track are termed as
``events of type C''. We have used different high $p_{t}$ cuts to check the
self shadowing effects in these event samples.

Figure 1(left) shows the plot of normalized variance ( $D^{2}$/$<$N$>$)
 of unbiased
multiplicity distribution versus centrality bin for Au+Au collisions at
$\sqrt{s_{NN}} =$ 200 GeV. It is observed that normalized variance
decreases monotonically with increasing centrality. Similar trend is 
observed for Pb+Pb collisions at $\sqrt{s_{NN}} =$ 11 TeV (Fig.1
(right)). Non-monotonic decrease of normalized variance with increasing
centrality is not observed for both  Au+Au and Pb+Pb collisions. 
\begin{figure}
\includegraphics[scale=0.31]{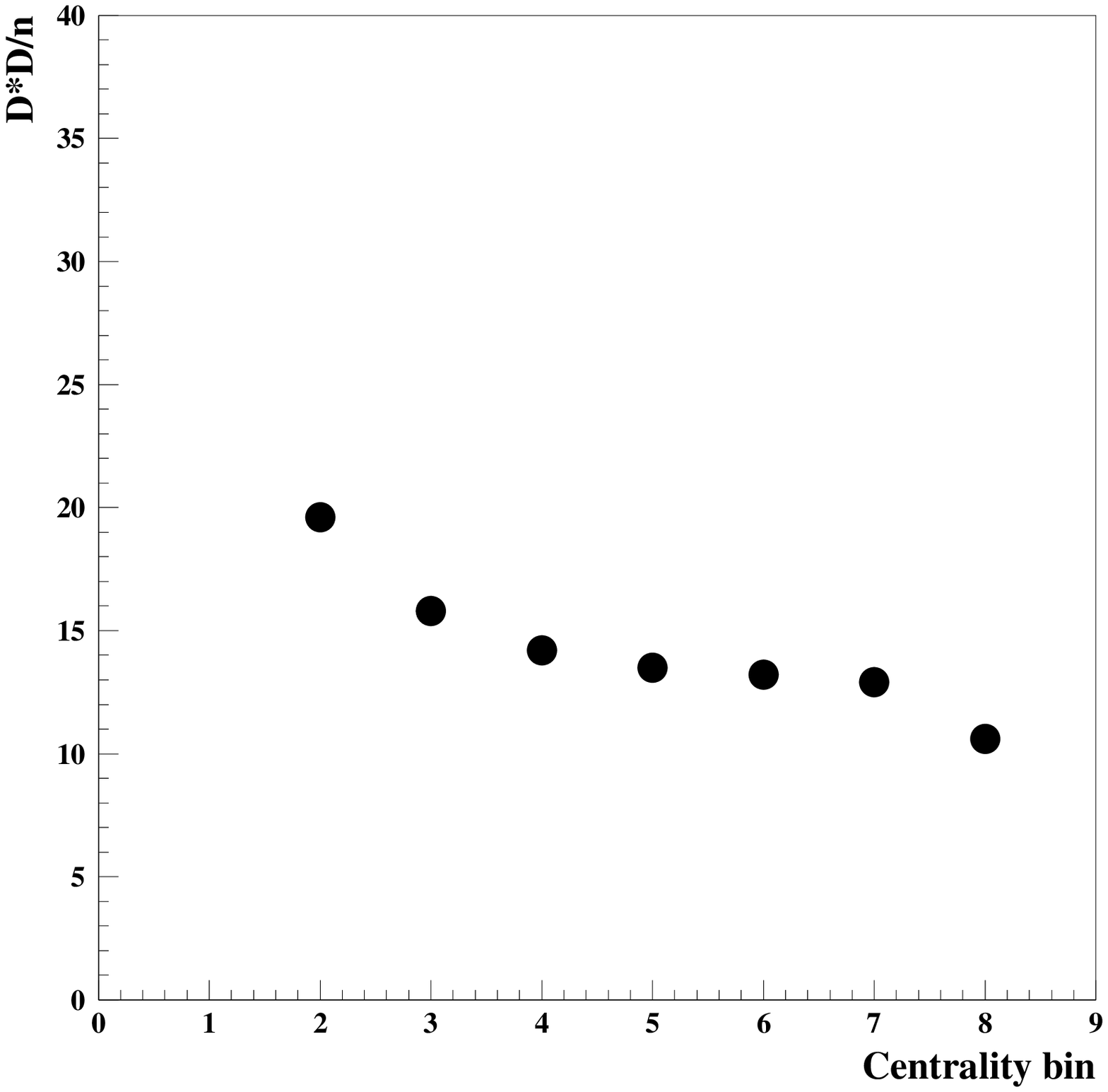}
\includegraphics[scale=0.31]{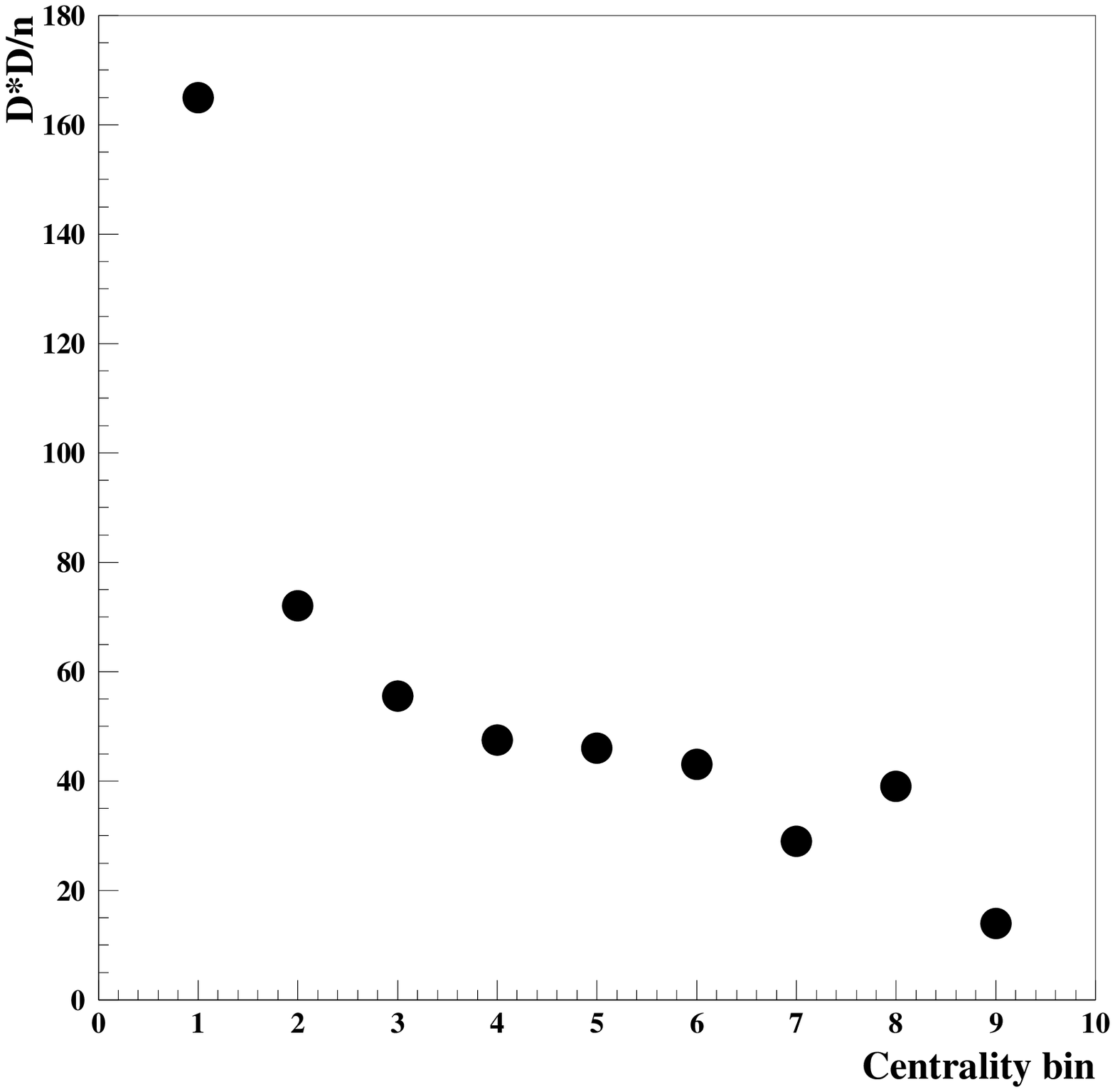}
\caption{Variation of normalized dispersion with centrality for Au+Au at
200 GeV(left) and Pb+Pb at 11 TeV(right) }
\label{fig1}
\end{figure}

From Eq. 9, we expect the ratio(R)

\begin{eqnarray}
R = \frac{(<n>_{C}-<n>)<n>}{D^{2}}=1
\end{eqnarray}

In Fig. 2(left), the ratio, R, is plotted versus centrality bin for Au+Au 
data for different $p_{t}$ cuts i.e., $p_{t}$ $>$ 2 GeV/c, $p_{t}$ $>$ 3 GeV/c,
 $p_{t}$ $>$ 4 GeV/c and  $p_{t}$ $>$ 5 GeV/c. It is seen that for $p_{t}$ $<$ 4
GeV/c ratio decreases with increasing centrality whereas for $p_{t}$ 
$>$ 4 GeV/c it stays constant as expected from Eq. 10 for self-shadowing
of high $p_{t}$ events. Fig.~2(right) exhibits similar plots for Pb+Pb
collisions for  $p_{t}$ $>$ 6 GeV/c,$p_{t}$ $>$ 8 GeV/c, $p_{t}$ $>$ 10 GeV/c and
$p_{t}$ $>$ 12 GeV/c. Here also it is noticed that for $p_{t}$ $<$ 10 GeV/c
ratio decreases with increase in centrality but almost stays constant for
$p_{t}$ $>$ 10 GeV/c indicating thereby self shadowing effect for high
$p_{t}$ events. It is observed that $p_{t}$ cut changes with change in the
collision energy for observing self-shadowing effect. 
\begin{figure}
\includegraphics[scale=0.32]{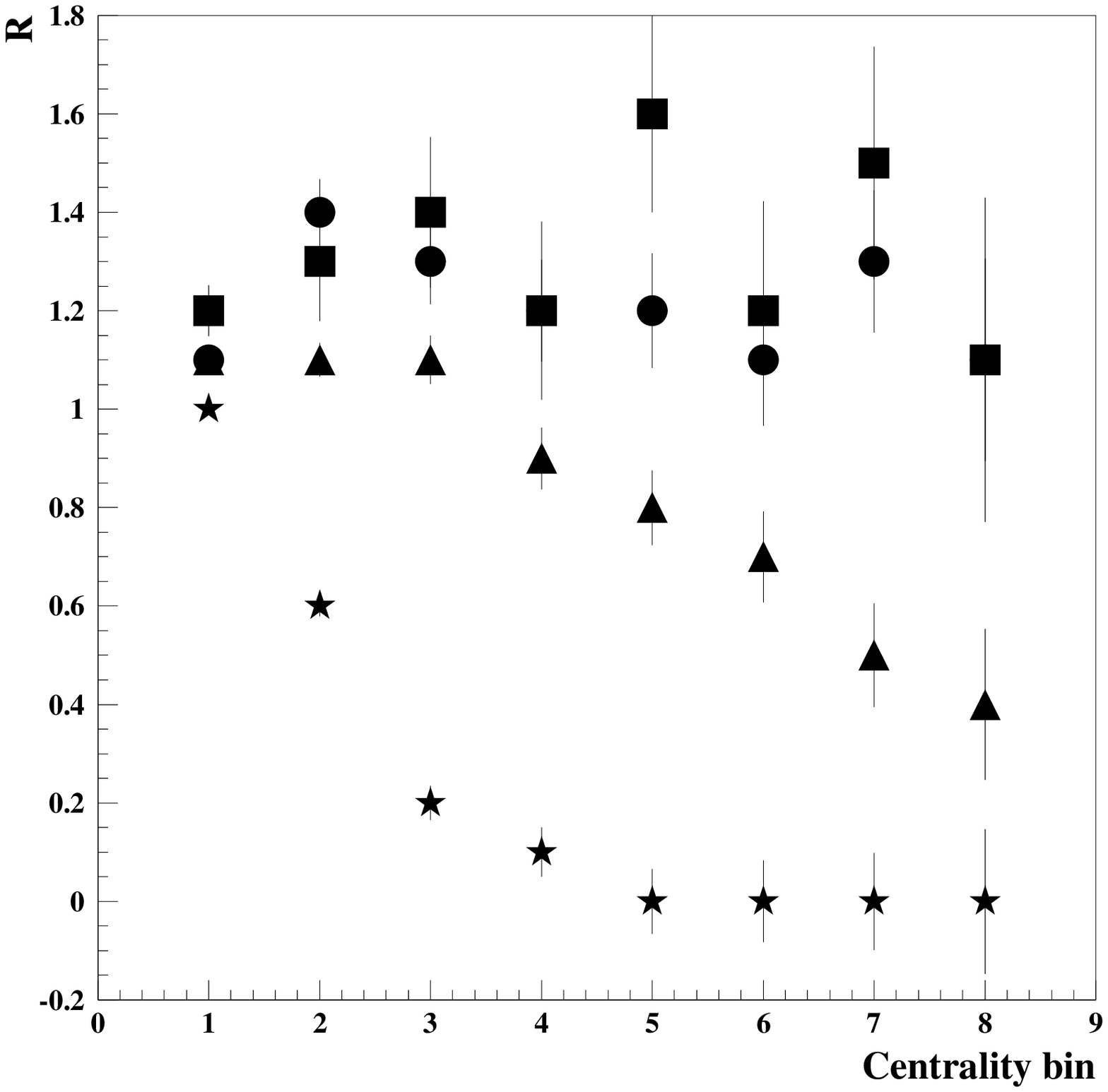}
\includegraphics[scale=0.32]{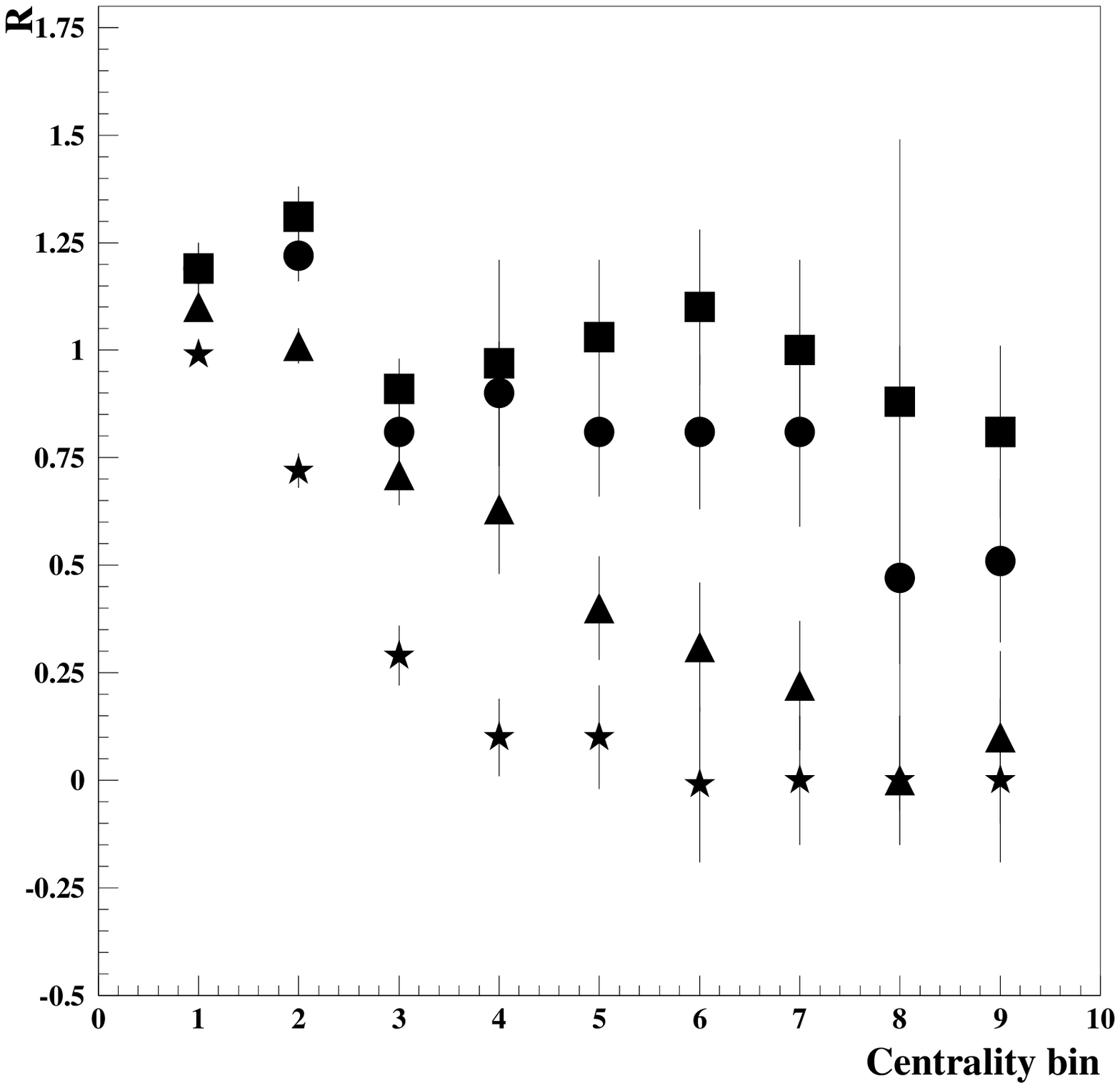}
\caption{Variation of R  with centrality for Au+Au at
200 GeV (left panel) and Pb+Pb at 11 TeV (right panel).
For left panel, results are shown for $p_{t}$ $>$ 2 GeV/c (stars),
 $p_{t}$ $>$ 3 GeV/c (triangles), $p_{t}$ $>$ 4 GeV/c (circles), and
$p_{t}$ $>$ 5 GeV/c (squares).
Results in the right panel are for $p_{t}$ $>$ 6 GeV/c (stars), $p_{t}$ $>$ 8 GeV/c (triangles), $p_{t}$ $>$ 
10 GeV/c (circles), and $p_{t}$ $>$ 12 GeV/c (squares).}
\label{fig2}
\end{figure}

An attempt has also been made to see if this trend is valid for photon
multiplicity distributions which can be observed with Photon Multiplicity
Detector(PMD) in STAR~\cite{ref6} at RHIC and in ALICE~\cite{ref7} at LHC. Here we have termed 
events as high $p_{t}$ if an event has at least one high $p_{t}$ charged
 particle
and studied the photon multiplicity distributions for different $p_{t}$
cuts on charged particles as PMD does not carry the information about
the momenta of photons. Fig.~3(left) displays the ratio for photons versus
centrality bin for Au+Au collisions. Here again we observed similar
trend as is seen in Fig. 2(left) for charged particles indicating that
self shadowing effect can be studied using photon multiplicity distributions
as well. In Fig. 3(right), we present the plot of ratio for photons
versus centrality bin for Pb+Pb collisions which again indicates that
high $p_{t}$ events are self shadowed.                                                                  
\begin{figure}
\includegraphics[scale=0.32]{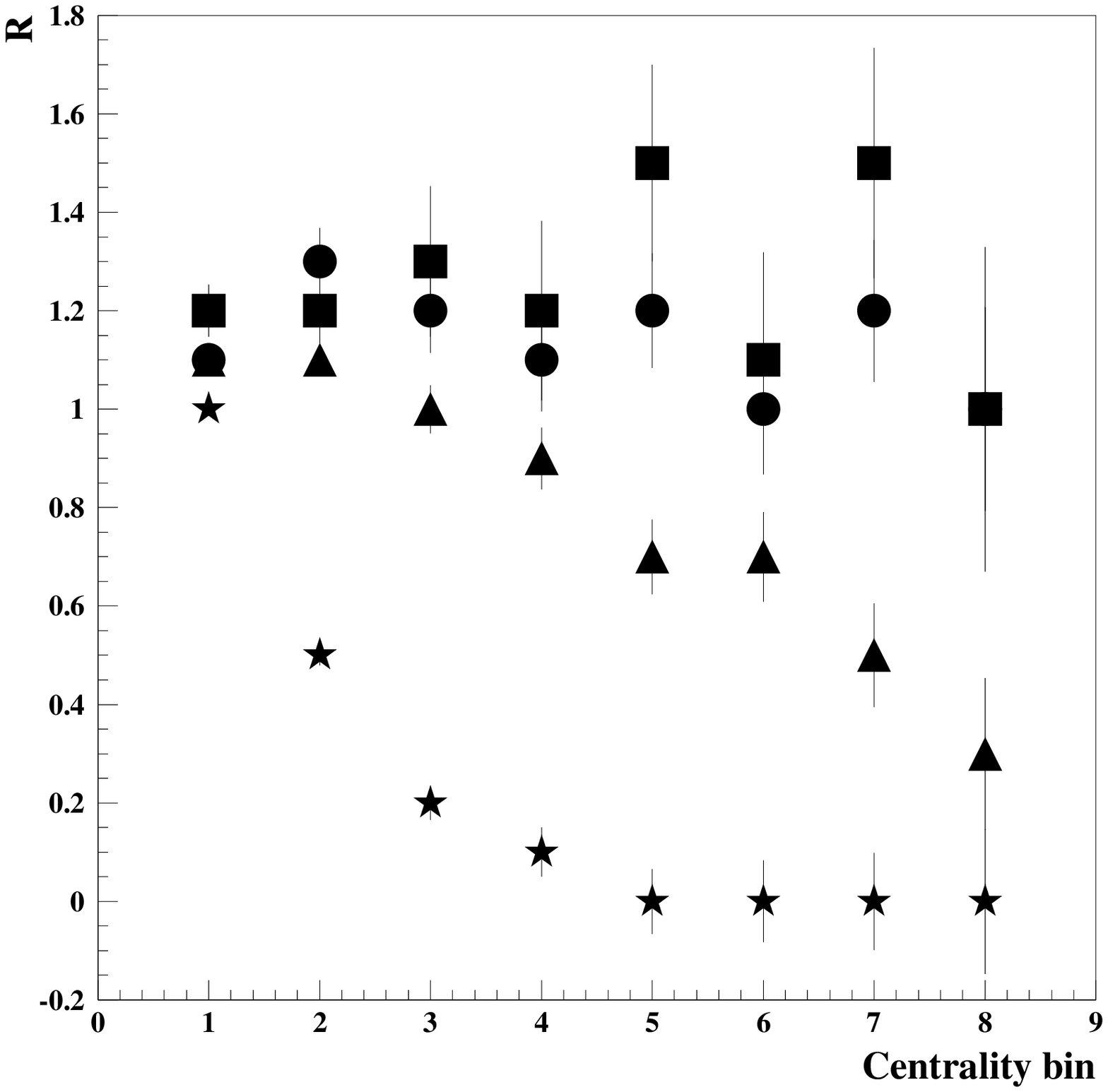}
\includegraphics[scale=0.32]{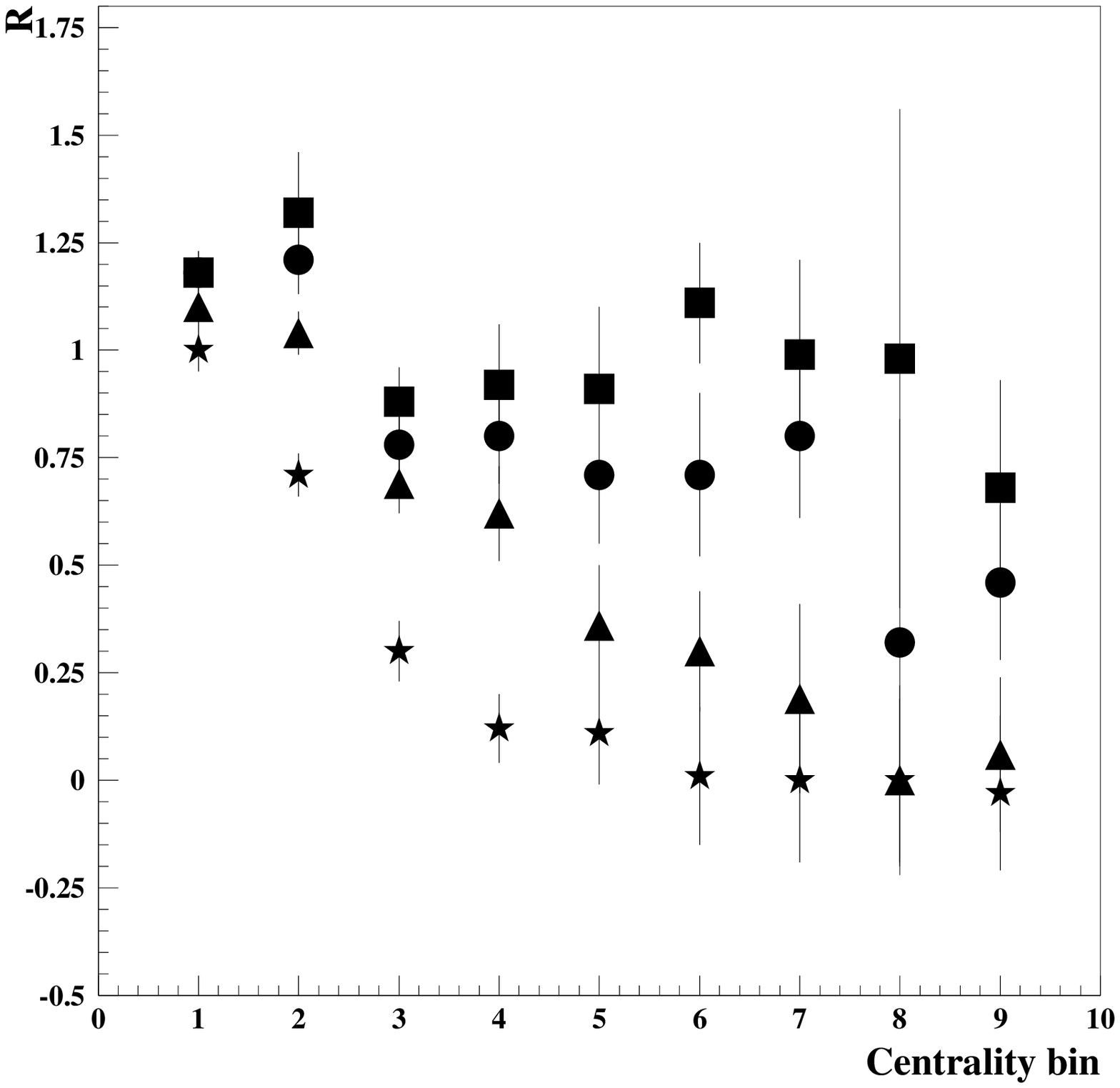}
\caption{Same as in Fig. 2 above but for photons.}
\label{fig3}
\end{figure}

\section{Summary}
Au+Au at $\sqrt{s_{NN}}$=200 GeV and Pb+Pb at $\sqrt{s_{NN}}$=11 TeV Hijing events exhibits self 
shadowing effect for high p$_{t}$ events.
It is observed that $p_{t}$ cut changes with change in the
collision energy for observing self-shadowing effect.
Photon multiplicity distribution also shows similar self-shadowing for high p$_{t}$ events. 
This can be checked using Photon Multiplicity Detector in STAR at RHIC and ALICE at LHC.\\


\begin{thebibliography}{50}
\medskip
\bibitem{ref1} K. Adcox et al. (PHENIX Collaboration), Nucl. Phys. A 757 (2005) 184, 
Phys. Rev. C 66 (2002) 024901; J. Adams et al.,(STAR Collaboration), Nucl. Phys. A 757 (2005) 102.
\bibitem{ref2} L. Cunqueiro, J. Dias de Deus and C. Pajares, Phys. Rev. C 74
(2006) 034901.
\bibitem{ref3} C. Pajares and A. V. Ramallo, Phys. Lett. B 107 (1981) 106.
\bibitem{ref4} J. Dias de Deus, C. Pajares, and C. A. Salgado, Phys. Rev. B 408 (1997) 417.
\bibitem{ref5} C. Pajares and A. V. Ramallo, Phys. Rev. D 31 (1985) 2800.
\bibitem{ref6} J. Adams et al.,(STAR Collaboration) Phys. Rev. Lett. 95 (2005) 062301.
\bibitem{ref7} ALICE PMD, Technical Design Report, CERN/LHCC 99-32(1999).
\end{thebibliography}
\end{document}